\documentstyle[12pt]{article}
\begin{document}
\title{Hard Diffraction From Quasi-Elastic Dipole Scattering}
\author{A.Bialas \\Institute of Physics, Jagellonian
University\\ Reymonta 4,  30-059 Cracow, Poland\\ and \\ 
Laboratoire de Physique Theorique et Hautes Energies\\
 Bat 211, Universite
Paris XI,
 F-91405 Orsay, France\thanks{Laboratoire associe au Centre National de
la Recherche Scientifique - URA D0063} \\ \\
R.Peschanski\\CEA, Service de Physique
Theorique, CE-Saclay\\ F-91191 Gif-sur-Yvette Cedex, France}
\maketitle
\begin{abstract}
The contribution to  diffraction dissociation of  virtual photons due to quasi-elastic scattering of the $q$-$\bar q$ component  is calculated in the framework of the QCD dipole picture. Both longitudinal and transverse components of the cross-section are given. It is shown that, at fixed mass of the diffractively produced system, quantum mechanical interference plays an important r\^ ole. Phenomenological consequences are discussed.
\end{abstract}

 {\bf 1.} Recently, new measurements of the proton diffractive structure function at
small $x$ and  large $Q^2$ were presented by H1 and ZEUS experiments
\cite{h1,zeus}.  The observed   3-dimensional
structure function:
\begin{equation}
F_2^{D(3)}(x_{Bj},Q^2,x_{\cal P}) = \frac {Q^2} {4\pi^2 \alpha_{e.m.}}
\frac{\beta }{x_{\cal P}}\frac{d\sigma} {d\beta}  = \frac {Q^4} {4\pi^2 \alpha_{e.m.}}
\frac 1{\beta x_{\cal P}}\frac{d\sigma} {dM^2}
\label{F3}
\end{equation}
can be studied as a function of $\beta =  Q^2/
(Q^2 + M^2),$ $x_{\cal P} = x_{Bj}/\beta,$ and
$M^2,$  the mass of the diffractively excited system.

The aim of the  present paper is to investigate the  QCD
contribution to the process in question using the colour
 dipole approach \cite{muel,nik}, which is known \cite{muel,nik,muela,muelb} to
reproduce the physics
of the "Hard Pomeron"\cite{lip}. We calculate the diffraction dissociation of
the virtual photon on the proton  at small $x_{Bj}$ and large enough $Q^2.$
Our motivation is two-fold: perturbative QCD calculations performed in a  framework suitable
 for  small $x_{Bj}$, such as the QCD dipole model, can provide a hint for discussing the physical mechanisms (perturbative vs. non-perturbative) at work in hard diffraction. Moreover, the same exercise performed for the structure function of the proton (which contains also the hard-diffraction contribution) leads \cite{NPR} to a correct description of the data with a non-perturbative input  restricted to some dipole-model parameters. We are thus looking for a  description of hard diffraction in the same framework.

In a first paper \cite{BP1}, we have derived the {\it inelastic} dipole-dipole cross-section contribution to the single diffraction dissociation of the
photon, see Fig(1-a). This component is expected to be dominant in the small-$\beta$ region where it corresponds to the triple-pomeron coupling in Regge language. However a {\it quasi-elastic} dipole-dipole collision is also present
(see Fig.(1-b)) and, moreover, is expected to dominate the finite $\beta$ domain. The investigation of this component is compulsory, since most of the present experimental results come from that region. 

This {\it quasi-elastic} dipole-dipole component has to be calculated for all the polarizations of the incident virtual photon, i.e. transverse and longitudinal ones. At  $\beta \simeq 1$  indeed, the longitudinal component is expected to dominate \cite{strik}. In the  present paper, we derive the predictions for cross-sections of these transverse and longitudinal photon components in the QCD dipole formulation. 
 Together with the results of \cite{BP1}, it provides a consistent set of   predictions for hard 
single-diffractive processes studied at HERA.

 The organization of the paper is the following: In the next section  we show how to derive the general 
 formulae for the diffractive cross-section of a virtual photon. The specific case of a hard diffractive cross-section in the QCD dipole picture is treated in section 3. In section $4$ we  discuss the implications of our calculations for
 the 3-dimensional diffractive structure function  at  small $x_{\cal P}.$ Our results are summarized in the final section.

{\bf 2.} The amplitude
for producing a mass $M$ in the process depicted in Fig. (1-b) reads
\begin{equation}
< M^2 k \mid T \mid \Psi_Q >=\int dz \frac{d^2\rho}{2\pi} <M^2 k \mid z k>
e^{ik\rho} T(\rho,b;x_{\cal P}) <z,\rho\mid\Psi_Q>  \label{mat}
\end{equation}
where $T(\rho,b;x_{\cal P})$ stands for the elastic amplitude for scattering of the
dipole with transverse size $\rho$ off the proton 
at the impact parameter $b$, $z$ is the light-cone momentum fraction of 
one of the quarks forming the dipole,  $k$ is the relative transverse momentum of the quark and
antiquark forming the dipole and $<z,\rho\mid\Psi_Q>$ is the probability amplitude
to find the dipole inside the photon. For fixed color, flavor and spin configuration of the $q \bar q$ pair $<z,\rho\mid\Psi_Q>$
is given by \cite{BKS}:
\begin{equation}
 <z,\rho|\Psi_Q> = \frac {\sqrt {\alpha_{e.m.}} e_{(f)}}{2\pi}\ \ \ z \hat Q e^{+ i\Phi} K_1 (\hat Q \rho)\label{BKS+}
\end{equation}
\begin{equation}
 <z,\rho|\Psi_Q> = \frac {\sqrt {\alpha_{e.m.}} e_{(f)}}{2\pi}\ (1-z) \hat Q e^{- i\Phi} K_1 (\hat Q \rho)\label{BKS-}
\end{equation}
\begin{equation}
 <z,\rho|\Psi_Q> = \frac {\sqrt {\alpha_{e.m.}} e_{(f)}}{2\pi}\ 2 z (1-z)  \ Q \ K_0 (\hat Q \rho) \label{BKSL}
\end{equation}
for the right handed, left handed and  longitudinal photons, respectively.
Here $\alpha_{e.m.}$ is the fine coupling constant, and $e_{(f)}$ is the quark charge and $\hat Q \equiv \left(z(1-z)\right)^{\frac12} Q $ (quark masses are neglected).

The corresponding cross-section (for each  component separately) reads
\begin{equation}
\frac {d\sigma}{dM^2 d^2b} = \int d^2k \mid <M^2,k\mid T \mid \Psi_Q> \mid ^2.
\label{sigen}
\end{equation}
Using the identity 
\begin{equation}
<M^2,k\mid z,k> =  \delta \left(M^2 - \frac{k^2}{z(1-z)}\right)
\frac{\left| 1-2z\right|}{z(1-z)}^{1/2}\ k   \label{mz}
\end{equation}
and introducing (\ref{mat}) into formula (\ref{sigen}), one can take advantage
of the two delta functions to  perform the
integration over k and one of the integrations over z. The result is:

\begin{equation}
\frac{d\sigma}{dM^2 d^2b} = \pi
\int_0^1 dz \ z(1-z)\ \left| G(\hat M, z;x_{\cal P})\right|^2
\label{sigd}
\end{equation}
where $\hat M \equiv \left(z(1-z)\right)^{\frac12} M $ and $G(k, z;x_{\cal P})$ is given by the  Fourier Transform:
\begin{equation}
 G(k,z;x_{\cal P}) = \frac 1{2\pi}\int d^2\rho \ e^{i \underline k . \underline {\rho}}\ 
T(\rho,b;x_{\cal P}) <z,\rho|\Psi_Q> .
\label{G}
\end{equation}

Eqns. (\ref {sigd},\ref {G}) show that the cross-section $d\sigma/dM^2 d^2b$ cannot be represented as a sum of interaction probabilities of dipoles with different sizes. It requires instead the use of the genuine BKS \cite {BKS} wave-functions (\ref {BKS+} - \ref{BKSL}). This is the consequence of the
fact that the mass $M$ of the diffractive remnants of the photon is 
not a "good" quantum number in the basis of states $|z,\rho>,$ known \cite{muel,nik} to be diagonal with respect to  dipole interactions. Thus, different dipole quantum states are contributing to the cross-section. This is the reason why we called the corresponding component {\it quasi}-elastic. Only when  integrated upon the mass $M,$  can the cross-section be represented by a sum of probabilities \cite{good}.

Inserting expressions (\ref{BKS+}-\ref{BKSL}) into (\ref{G}), one can perform the integration over azimuthal angle and one obtains:
\begin{equation}
G(z,k,b;x_{\cal P}) =  \frac {\sqrt {\alpha_{e.m.}} e_{(f)}}{2\pi}\ \ z \hat{Q} \int \rho d\rho T(\rho , b;x_{\cal P}) K_1(\hat{Q}\rho)
J_1(k\rho),   \label{G+}
\end{equation}
\begin{equation}
G(z,k,b;x_{\cal P}) =   \frac {\sqrt {\alpha_{e.m.}} e_{(f)}}{2\pi}\ (1-z )\hat{Q} \int \rho d\rho T(\rho , b;x_{\cal P}) K_1(\hat{Q}\rho)
J_1(k\rho),   \label{G-}
\end{equation}
\begin{equation}
G(z,k,b;x_{\cal P}) = \frac {\sqrt {\alpha_{e.m.}} e_{(f)}}{2\pi}\ 2z(1-z)\ Q \int \rho d\rho T(\rho , b;x_{\cal P}) K_0(\hat{Q}\rho)
J_0(k\rho).   \label{GL}
\end{equation}

{\bf 3.}
The formulae of section 2 are general and can be used for any model of the interaction amplitude $T$ of a dipole on  any target.  In the QCD dipole picture, and if the target is an onium,  $T$ was calculated \cite{muel} and is given by:

\begin{equation}
T(\rho,b;x_{\cal P}) = \int d^2r dz\ \Phi (r,z) \ \tilde T(r,\rho,b;x_{\cal P}), 
\label{tT}
\end{equation}
where,  for $b> max(r,\rho),$

\begin{equation}
\tilde T(r,\rho,b;x_{\cal P}) = \pi \alpha^2 \frac{r \rho}{b^2} \log \left(\frac{b^2}{r\rho}
\right) x_{\cal P}^{-\Delta_{\cal P}} (2a_{\cal P}/\pi)^{3/2}\exp\left(-\frac{a_{\cal P}}{2}
\log^2(b^2/r\rho) \right ).   \label{fT}
\end{equation}
$\Phi (r,z)$ is the square of the onium wave function, $\alpha$ is the strong coupling constant, $\Delta_{\cal P} \equiv \alpha_{\cal P} -1 =
\frac{\alpha N_c}{\pi} 4 \log 2\ ,$ and 

 \begin{equation}
    a_{\cal P} = [7\alpha N_c\zeta(3)\log (1/x_{\cal P})/\pi]^{-1}.      \label{aP}
\end{equation}
In the limit $x_{\cal P} \rightarrow 0$ the integral (\ref{tT}) can be approximated \cite{BP1} by:
 \begin{equation}
T(\rho,b;x_{\cal P}) = \tilde T(r = r_0,\rho,b;x_{\cal P}),
\label{approx}
\end{equation}
 where 
 \begin{equation}
r_0 = \int d^2r\ dz \ r\ \Phi (r,z),
\label{r0}
\end{equation}
and  $b> max(r_0,\rho).$ 

To apply these results to deep inelastic lepton-nucleon scattering,  one makes the assumption that the target proton configurations contributing at small $x_{Bj}$ can be adequately represented by onia. The same assumption was shown to work well for the proton structure function itself \cite{NPR}. In that case $r_0$ is a non-perturbative parameter which is determined from the data. The fit to the data on proton structure function gives
$r_0 \simeq .8 $ fermi.

Inserting (\ref{tT},\ref{fT}) into (\ref{G+},\ref{G-},\ref{GL}) and (\ref{sigd}), and summing over all color, flavor and spins, 
one obtains an explicit expression for the diffractive cross-section we are looking for, namely
\begin{equation}
\frac {d\sigma}{dM^2} = \frac {d\sigma_T}{dM^2} +  \frac {d\sigma_L
}{dM^2}
\label{1}
\end{equation}
with
\begin{eqnarray}
 \frac {d\sigma_T}{dM^2} = \frac {N_c \alpha_{e.m.} e^2_f}{2\pi}\ Q^2\ \int_{r_0}^\infty d^2 b \int_0^1 dz \ z^2(1-z)^2 \left(z^2 + (1-z)^2\right)\times \nonumber \\ \times \left| \int_0^b \rho d\rho T(\rho , b;x_{\cal P}) K_1(\hat{Q}\rho)
J_1(\hat M\rho) \right|^2 
\label{2}
\end{eqnarray}
\begin{eqnarray}
 \frac {d\sigma_L}{dM^2} = 4\ \frac {N_c \alpha_{e.m.} e^2_f}{2\pi}\ Q^2\ \int_{r_0}^\infty d^2 b \int_0^1 dz \ z^3(1-z)^3 \times \nonumber \\ \ \ \ \ \times \left| \int_0^b \rho d\rho T(\rho , b;x_{\cal P}) K_0(\hat{Q}\rho)
J_0(\hat M\rho) \right|^2 ,
\label{3}
\end{eqnarray}
where $e^2_f = \sum e_{(f)}^2,$ is the sum of the square charges of the quarks.

The formulae (\ref{1}-\ref{3}) involve an integration on the dipole size $\rho$
from $0$ to $b.$ As $b$ can take rather large values, the question arises whether the non-perturbative effects do not modify significantly  the results \cite{Bj}. In order to discuss this problem, we investigated the dependence of the results on a cut-off $r^* $ of integration in $\rho,$ i.e. we replaced the integration from $0$ to $b$ by $0$ to min($r^\ast,b$). The results are illustrated in Fig. 2 where the dependence of $d\sigma_T/dM^2$ and $d\sigma_L/dM^2$ on $r^\ast$ is plotted.
One observes that for the longitudinal component there is no variation for $r^\ast > r_0.$ For the transverse component there is little variation at small  $\beta$ and somewhat more at larger values of $\beta.$  We also checked that these results are not sensitive to $M^2 + Q^2,$ that is to the dimensionful scale of the problem. It means that the process in question is fully controlled by short-distance physics only if $r_0$ is small enough. The case of proton target, with $r_0 \simeq .8f,$ does not satisfy this condition \cite{nik,Bj}. On the other hand, since $r_0$ is not unreasonably large, one may hope  that the corrections 
to the photon wave functions (\ref{BKS+}-\ref {BKSL}), expected from non-perturbative (long
distance) forces are not very important. Since we  find that the bulk of the cross-section comes from the region $\rho <  r_0, $ from now on we consider the results for $r^{\ast} = r_0.$

{\bf 4.} Let us now discuss the main phenomenological features of this {\it quasi-elastic} component of the 3-dimensional diffractive structure function (1). 

i) $x_{\cal P}$-dependence. 

One sees from formula (\ref{tT}) that the factor
\begin{equation}
\Phi_{\cal P} =  \  x_{\cal P}^{-1-2\Delta_{\cal P}}
\left(\frac{2a_{\cal P}}
{\pi} \right)^3 \label{flux}
\end {equation}
common to all components, is responsible for most of the $x_{\cal P}$-dependence. However an extra, a-priori non-factorizable, contribution is also present.  It comes from  the factor $ \exp\left(-\frac{a_{\cal P}}{2}
\log^2(b^2/r\rho) \right )$ in  (\ref {fT}). Its effects are displayed in Fig. 3 where
\begin{equation}
 \Delta_{\cal P}^{eff} = \frac12\ \left( \frac {d\log F^{D(3)}}{d\log (1/x_{\cal P})} - 1 \right)
 \label{eff}
\end{equation}
is plotted versus $x_{\cal P}.$ One sees that the predicted effective intercept
is higher than the one found for triple-pomeron contribution \cite{BP1} (coming from $\Phi _{\cal P}$ only). Nevertheless, it is still substantially lower than $\Delta_{\cal P},$ and from the effective intercept determined from the total cross-section \cite {AB}. Some violation of factorization is present because $ \Delta_{\cal P}^{eff} $ depends on $Q^2.$ These effects are rather small however, and it will be difficult to observe in the data. We have  checked that the same is even more true for the $\beta$ dependence of $ \Delta_{\cal P}^{eff}. $ So, the {\it quasi-elastic} component appears to be effectively factorizable.

ii) $\beta$-dependence.

It is shown in Fig. 4 for $Q^2 = 8.5$ and $x_{\cal P} = 10^{-3}.$ It is remarkable that, although the  transverse and longitudinal components show very different behaviours, the sum is almost independent of $\beta$ for $\beta > .3$. The dominance of the longitudinal over the transverse contribution when 
$\beta \rightarrow 1$ is a direct consequence of the vanishing of $J_1$ in the integral (19). The noticeable consequence is that the ratio $R$ of the longitudinal over the transverse component is predicted to become larger than 1 for $\beta > .8.$

iii) $Q^2$-dependence.
 The structure function $F^{D(3)}$ was found to increase with $Q^2$
in the whole range of $\beta$, in contrast to the situation 
encountered in the proton structure function $F_2$. This seems in 
qualitative agreement with the data \cite {h1}. This effect, however, is
sensitive to the scales involved in the process and thus requires more
careful analysis before definite conclusion can be reached.

{\bf 5.} To summarize, we have calculated the contribution to diffractive
dissociation of a virtual photon at small $x$ due to quasi-elastic
scattering of the $q\bar{q}$ component of the photon on the target
nucleon (Fig.1). The amplitude for dipole-proton scattering was
taken to be that corresponding to the exchange of a "hard Pomeron",
as calculated \cite{muel} in the dipole picture of high-energy
scattering \cite{muel,nik}  (as shown
in \cite{NPR} it gives a good description of the proton structure
function $F_2$ at small $x$).  Our main results can be formulated as
follows:

(a) At a fixed mass of the diffractively produced system, the cross-
section for the process in question cannot be represented as a sum 
of the interaction probabilities of different dipole components of the virtual
photon. The quantum-mechanical interference is essential and thus the
explicit use of the light-cone wave functions of the virtual photon
 \cite{BKS} is necessary, independently of the model describing the dipole target elastic amplitude. 

(b)   We found that the main part of the cross-section is coming from 
the interaction of the $q\bar{q} $ pairs whose transverse size is of the order of the target size  as seen by the virtual photon. 
We found, furthermore, that this range 
does not depend on the dimensionful large scale
of the problem, i.e. $Q^2+M^2$. If the target size is of order  $1f,$ as for a proton, the process in 
question is not fully controlled by the short
distance physics \cite{nik,Bj}. On the other hand, since
the effective range is not unreasonably large, we feel that the corrections 
to the photon wave functions, expected from non-perturbative (long
distance) forces should not be overwhelming.
 
(c)  We find the effective Pomeron intercept 
(Fig.3) of the quasi-elastic
component significantly higher that that of the triple-Pomeron
contribution \cite{BP1} but still lower than that determined from
the proton structure function $F_2$ \cite{AB}, in  agreement
with  data \cite{h1,zeus}. Approximate factorization is found
with respect to both $\beta$ and $Q^2$ dependence, in agreement with the
results of \cite{nikital}.

(d) The obtained $\beta$-distribution (Fig.4) is fairly flat for $\beta
> .3$ because  the contribution from the longitudinal component fills the dip of the transverse one
at $\beta \approx 1  .$  Thus the ratio
R of the longitudinal to transverse diffractive structure functions
is expected to be strongly varying as function of $\beta$ and to reach
values exceeding $1$ for $\beta$ greater than $0.8$.

(e) The structure function $F^{D(3)}$ was found to increase with $Q^2$
in the whole range of $\beta$, in contrast to the situation 
encountered in the proton structure function $F_2$.

In conclusion, the present calculation, together with the one of Ref.\cite {BP1}
show that the QCD dipole picture leads to a well-defined prediction for the hard diffractive process seen as large rapidity gap events at HERA. With the advent of new precise data, it will be possible to perform a quantitative comparison.

{\bf Acknowledgments}
Discussions
with A. Capella and Ch. Royon are appreciated. This work has been supported by
 the KBN grant No 2 P03B 083 08 and by PECO grant from
the EEC Programme "Human Capital and Mobility", Network "Physics at
High Energy Colliders", Contract Nr: ERBICIPDCT940613.
\

\eject

\bigskip
{\bf Figure captions.}
\bigskip
\bigskip

Fig.1. Graphs contributing to the diffraction dissociation of virtual photons.
(a) inelastic (triple-Pomeron)
(b) quasi-elastic
\bigskip

Fig.2. Dependence of the 3-dimensional diffractive structure function on the integration cut-off $r^*.$
\bigskip

Fig.3. Effective Pomeron intercept for triple Pomeron (dashed line) and quasi-elastic (full lines) contributions.
\bigskip

Fig.4. $\beta-$ dependence of the 3-dimensional structure function.
\eject

\input epsf
\vsize=25.truecm
\hsize=14.truecm

\epsfysize=10cm{\centerline{\epsfbox{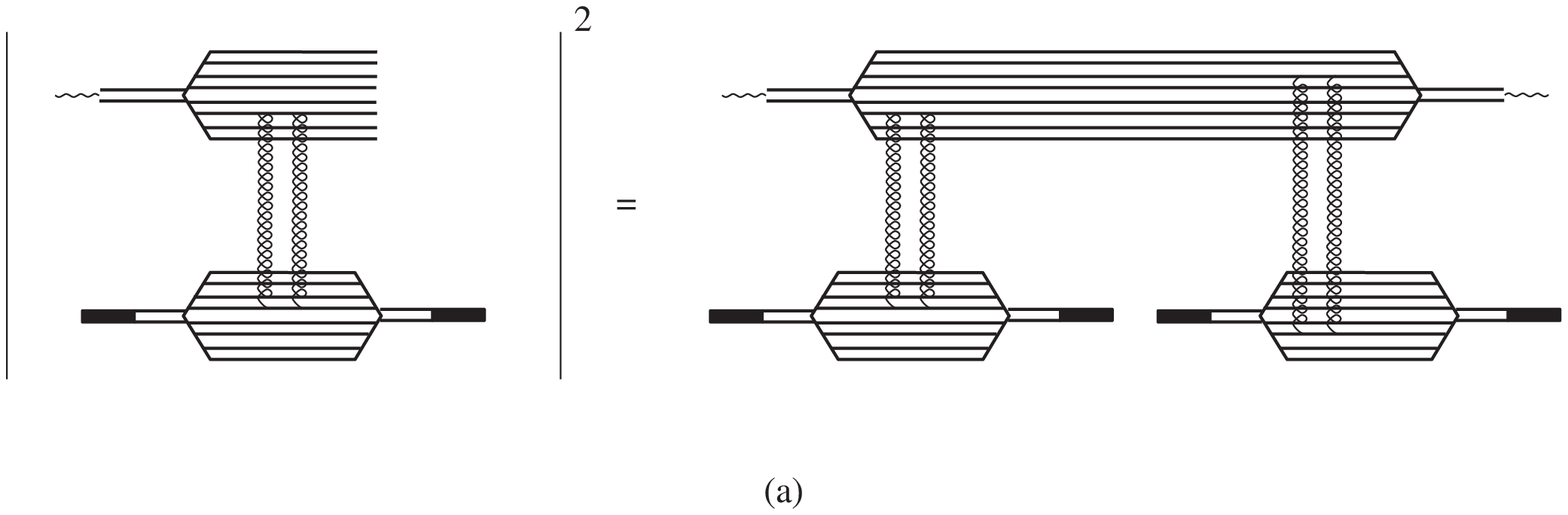}}}
\epsfysize=10cm{\centerline{\epsfbox{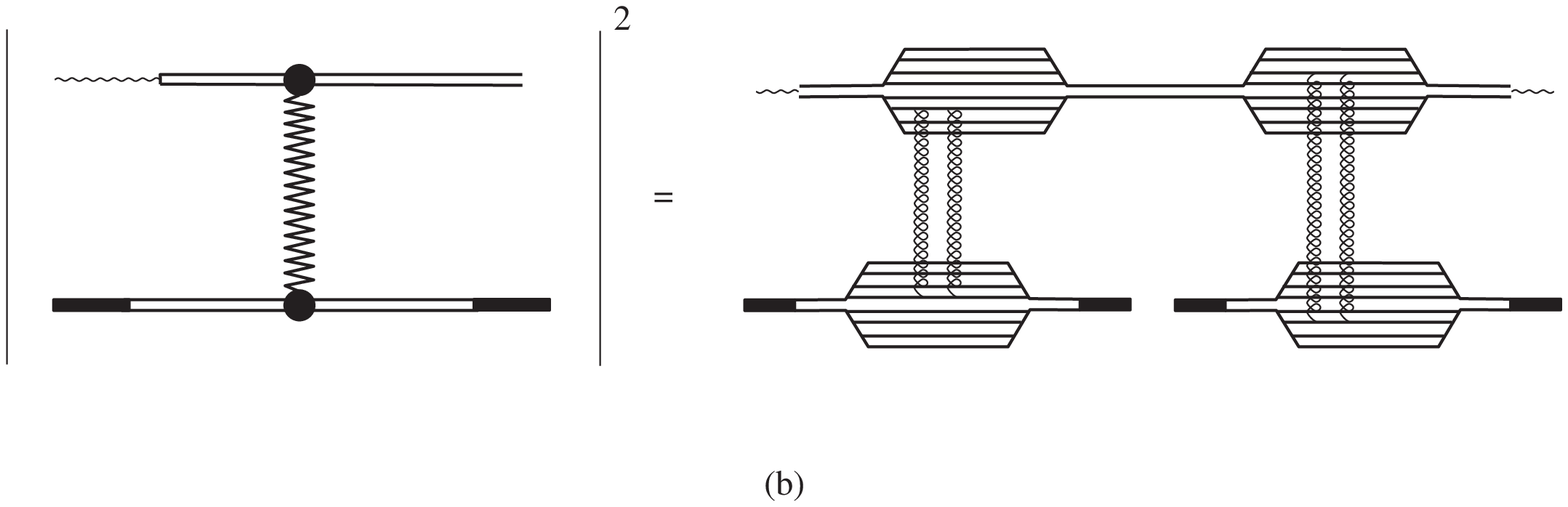}}}
\bigskip
\eject

%\vsize=25.truecm
%\hsize=18.truecm
\bigskip

\epsfysize=8cm{\centerline{\epsfbox{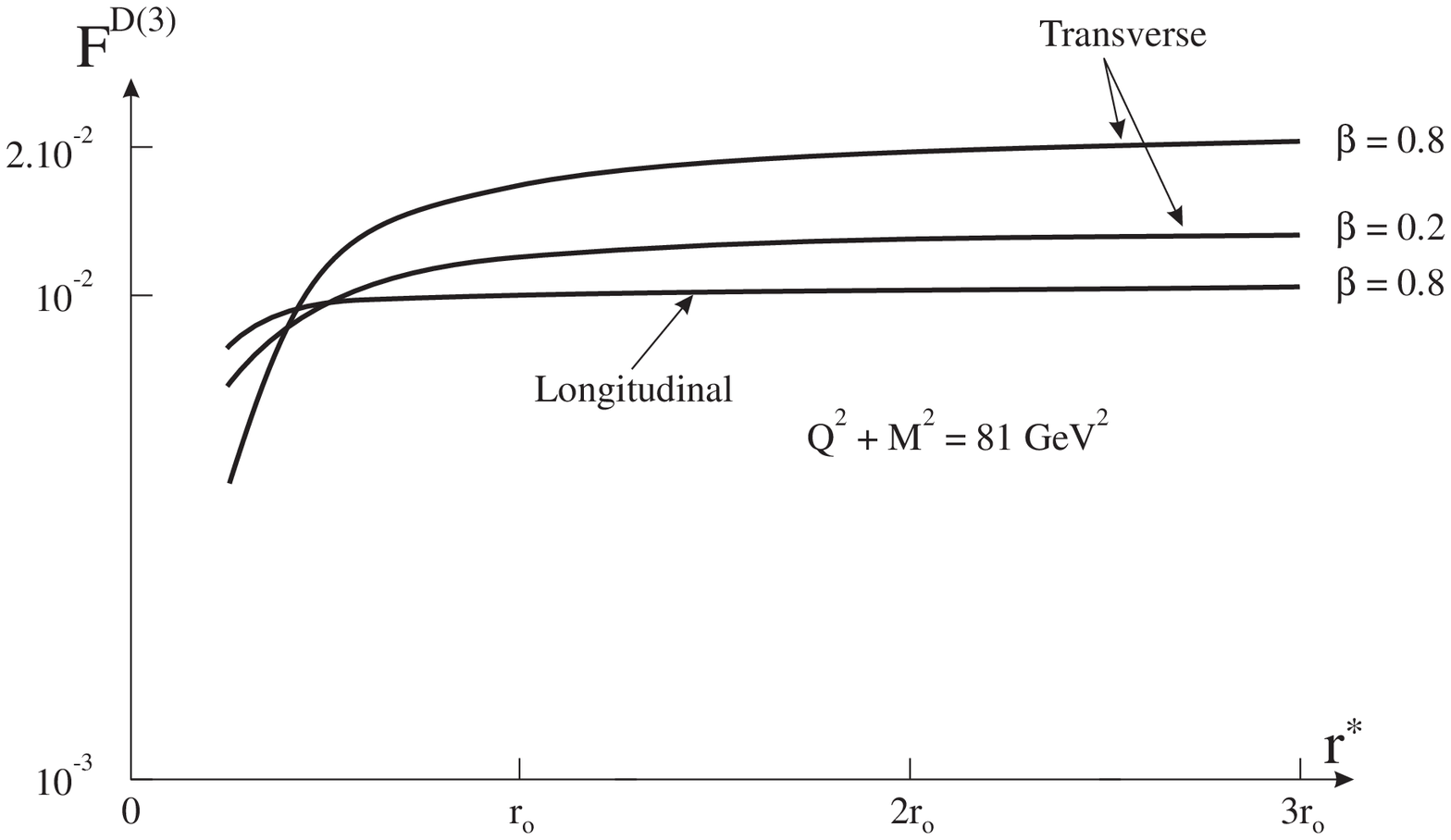}}}
\bigskip
\eject

\bigskip

%\vsize=25.truecm
%\hsize=18.truecm
\epsfysize=8cm{\centerline{\epsfbox{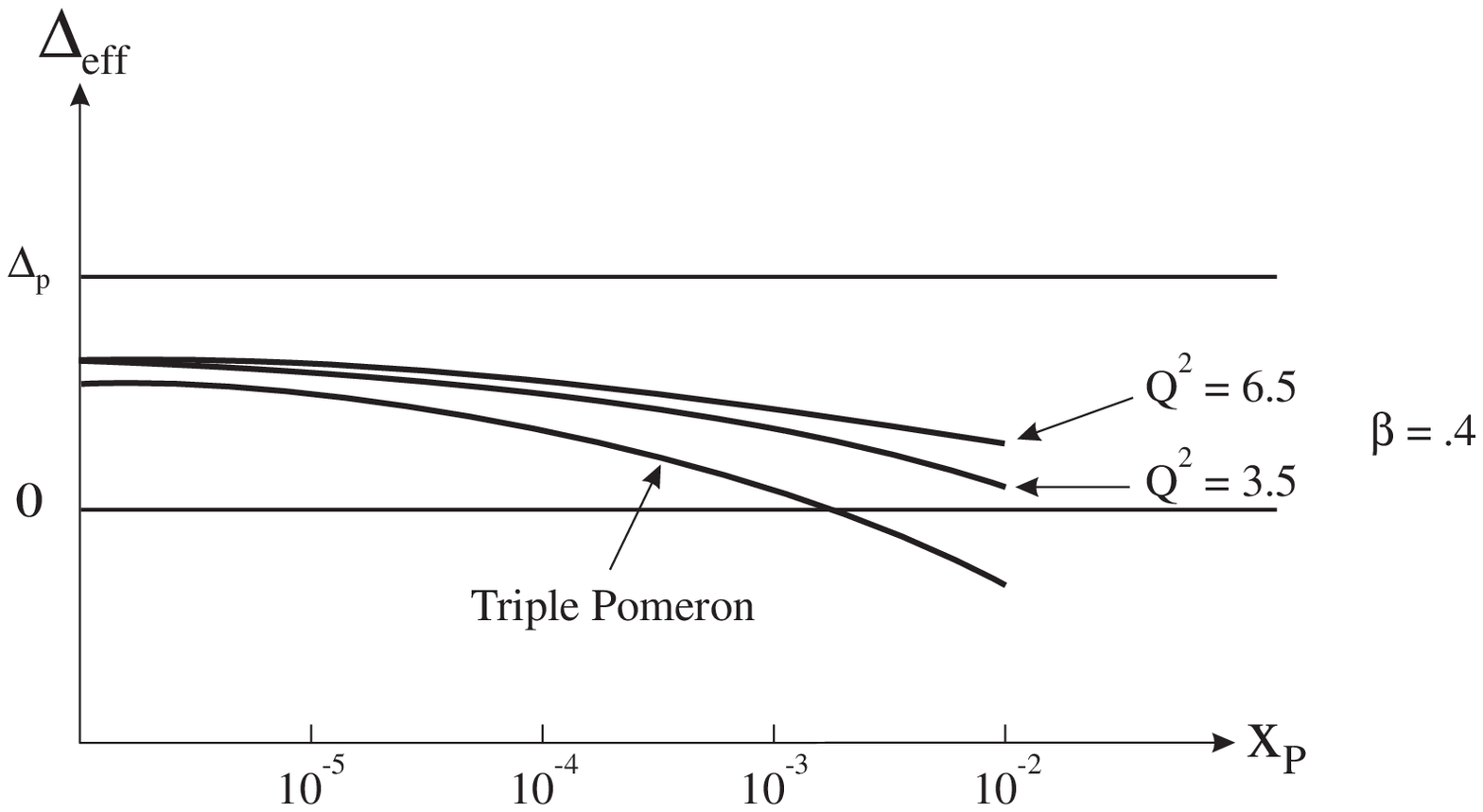}}}
\bigskip
\eject
\bigskip

%\vsize=25.truecm
%\hsize=18.truecm
\epsfysize=10cm{\centerline{\epsfbox{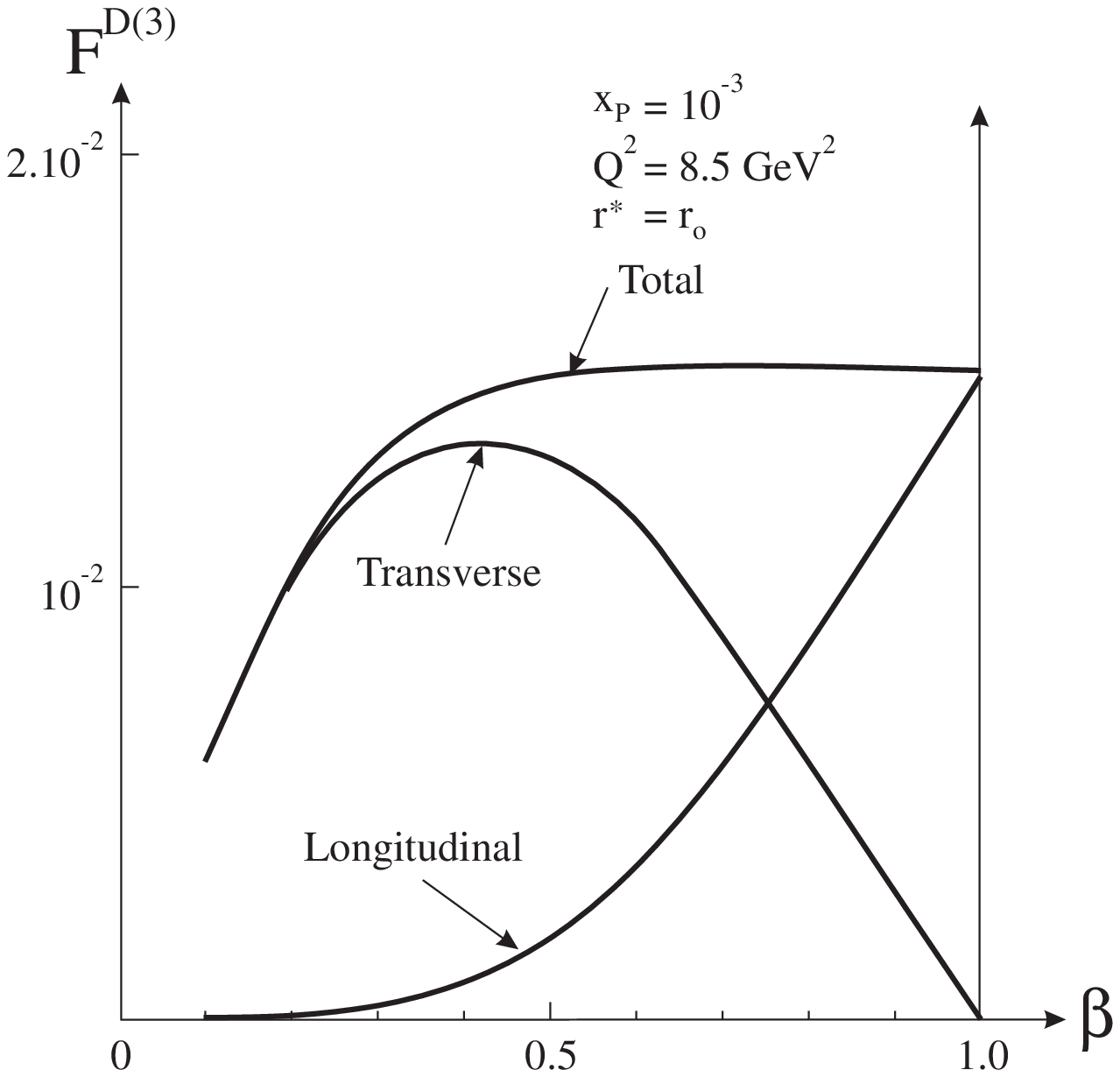}}}
\bigskip
\eject

\begin{thebibliography}{99}
\bibitem{h1}
H1 coll., T.Ahmed et al. \ {\it Phys.Letters} {\bf B348} (1995) 681.
\bibitem{zeus}
ZEUS coll., M.Derrick et al. {\it Zeit. fur. Phys.} {\bf C68} (1995) 569. \bibitem{muel}
A.H.Mueller, {\it Nucl. Phys.} {\bf B415} (1994) 373.
\bibitem{nik}
N.N.Nikolaev and B.G.Zakharov, {\it Zeit. fur. Phys.} {\bf C49} (1991) 607;
{\it ibid.} {\bf C64} (1994) 651.
\bibitem{muela}
A.H.Mueller and B.Patel, {\it Nucl. Phys.} {\bf B425} (1994) 471.
\bibitem{muelb}
A.H.Mueller, {Nucl. Phys.} {\bf B437} (1995) 107.
\bibitem{lip}
V.S.Fadin, E.A.Kuraev and L.N.Lipatov   {\it Phys. lett.} {\bf B60} (1975)
50; I.I.Balitsky and L.N.Lipatov, {\it Sov.J.Nucl.Phys.} {\bf 28} (1978) 822.
\bibitem{NPR}
H.Navelet,R.Peschanski and Ch.Royon, {\it Phys. Lett.} {\bf B366} (1995) 329.
\bibitem{BP1}
A. Bialas and R. Peschanski, {\it Hard Diffraction in the QCD Dipole Picture}, hep-ph/9512427, to be published in {\it Phys. Lett.} {\bf B}.
\bibitem{strik}
S. Brodsky et al. {\it Phys. Rev.} {\bf D50} (1994) 3144.
\bibitem{BKS}
J.Bjorken, J.Kogut and D. Soper, {\it Phys.Rev.} {\bf D3} (1971) 1382.
\bibitem{good} 
M.L.Good and W.D.Walker, {\it Phys. Rev.} {\bf 120} (1960) 1857.
H.I.Miettinen and J.Pumplin, {\it Phys.Rev.} {\bf D18} (1978) 1696.
\bibitem{Bj}
J.D.Bjorken, SLAC-PUB-7096, January 1996.
\bibitem{AB}
A.Bialas, {\it  BFKL Pomeron in the Impact Parameter Picture}
(Presented at the 2nd Cracow Epiphany Conference,  5-6
January 1996) To be published in Acta Physica Polonica B.
\bibitem{nikital}
M.Genovese, N.N.Nikolaev and B.G.Zakharov, Preprint KFA-IKP(Th)-1994-37 (unpublished).
\end{thebibliography}
\end{document}